\documentclass[sigplan,screen]{acmart}
\usepackage{hyperref}

\AtBeginDocument{%
  \providecommand\BibTeX{{%
    \normalfont B\kern-0.5em{\scshape i\kern-0.25em b}\kern-0.8em\TeX}}}

\usepackage{multirow}
\usepackage{array}
\begin{document}
\newcommand\MyBox[2]{
  \fbox{\lower0.75cm
    \vbox to 1.7cm{\vfil
      \hbox to 1.7cm{\hfil\parbox{1.4cm}{#1\\#2}\hfil}
      \vfil}%
  }%
}

\def\colorModel{hsb} 

\newcommand\ColCell[1]{
  \pgfmathparse{#1<50?1:0}  
    \ifnum\pgfmathresult=0\relax\color{white}\fi
  \pgfmathsetmacro\compA{0}      
  \pgfmathsetmacro\compB{#1/100} 
  \pgfmathsetmacro\compC{1}      
  \edef\x{\noexpand\centering\noexpand\cellcolor[\colorModel]{\compA,\compB,\compC}}\x #1
  } 
\newcolumntype{E}{>{\collectcell\ColCell}m{0.4cm}<{\endcollectcell}}  
\newcommand*\rot{\rotatebox{90}}

\settopmatter{printacmref=false}
\setcopyright{none}
\renewcommand\footnotetextcopyrightpermission[1]{}
\pagestyle{plain}

\pagestyle{fancy}
\lhead{Cross-Domain Consumer Review Analysis}
\chead{}
\rhead{Kunal Joshi, Aditya Pandey}
\cfoot{\thepage}

\title{Cross-Domain Consumer Review Analysis}


\author{Kunal Joshi}
\affiliation{%
  \institution{New York University}
  \city{New York}
  \country{USA}
}
\email{kkj7842@nyu.edu}

\author{Aditya Pandey}
\affiliation{%
  \institution{New York University}
  \city{New York}
  \country{USA}
}
\email{ap6624@nyu.edu}


\begin{abstract}

    The paper presents a cross domain review analysis on four popular review datasets: \href{https://www.amazon.com/}{Amazon}, \href{https://www.yelp.com/}{Yelp}, \href{https://store.steampowered.com/}{Steam}, and \href{https://www.imdb.com/}{IMDb}. The analysis is performed using Hadoop\cite{hdfs} and Spark\cite{spark}, which allows for efficient and scalable processing of large datasets. By examining close to 12 million reviews from these four online forums, we hope to uncover interesting trends in sales and customer sentiment over the years. Our analysis will include a study of the number of reviews and their distribution over time, as well as an examination of the relationship between various review attributes such as upvotes, creation time, rating, and sentiment. By comparing the reviews across different domains, we hope to gain insight into the factors that drive customer satisfaction and engagement in different product categories.
\end{abstract}


\settopmatter{printacmref=false}


\maketitle

\section{Introduction}
The increasing use of social media and online sales platforms has led to a proliferation of consumer reviews and other forms of digital feedback. Companies want to be customer-centric and understand their customers' needs and preferences in order to improve their products and services. By analyzing this digital footprint of customer opinions and behaviors, companies can gain valuable insight into the factors that drive customer satisfaction and demand. This information can help them develop effective marketing and advertising strategies that resonate with their customers and reach them through the channels they use. In this way, tapping into the digital footprint of customers can be a valuable tool for companies looking to build and grow their businesses. \cite{content_marketing}

Sentiment analysis, also known as opinion mining, is a field of study that examines people's attitudes and thoughts toward certain entities. This can be done by analyzing the content that users post on various social media platforms such as forums, microblogs, and online social networking sites. \cite{7336187} Many of these sites provide APIs that allow researchers and developers to collect and analyze the data. However, the quality of user-generated content can vary, and the availability of ground truth for the sentiments expressed may also be limited. A ground truth is a reference or standard against which the accuracy of sentiment can be measured. These factors can potentially hinder the process of sentiment analysis.

\section{Motivation}
Understanding customer behavior and needs are critical for companies, as it can help them evaluate customer satisfaction and make necessary improvements to their products and services. A centralized sentiment analysis system can provide more accurate and useful insights into customers' personal experiences, thoughts, and beliefs. This can be beneficial for businesses and advertising companies, as it can help them improve their offerings and increase profits. The users of such a system are typically business owners, data analysts, and data scientists. By analyzing customer sentiment, these individuals can help businesses and advertising companies make better decisions and more effectively target their audience. 

\section{Goodness}
We believe the results of our analytics are correct and can be trusted for the following reasons:
\begin{itemize}
    \item The results of the analysis were consistent with the sales of products on platforms such as Amazon and Steam
    \item It is observed that people have more free time on weekends, and are more likely to visit various businesses such as restaurants and movie theaters. This is reflected in the higher number of reviews posted on weekends.
    \item Many of the findings are supported by the expected behavior of users at different times.
\end{itemize}

\section{Datasets}
\label{sec:dataset}

\begin{figure*}[h!]
  \includegraphics[width=\textwidth]{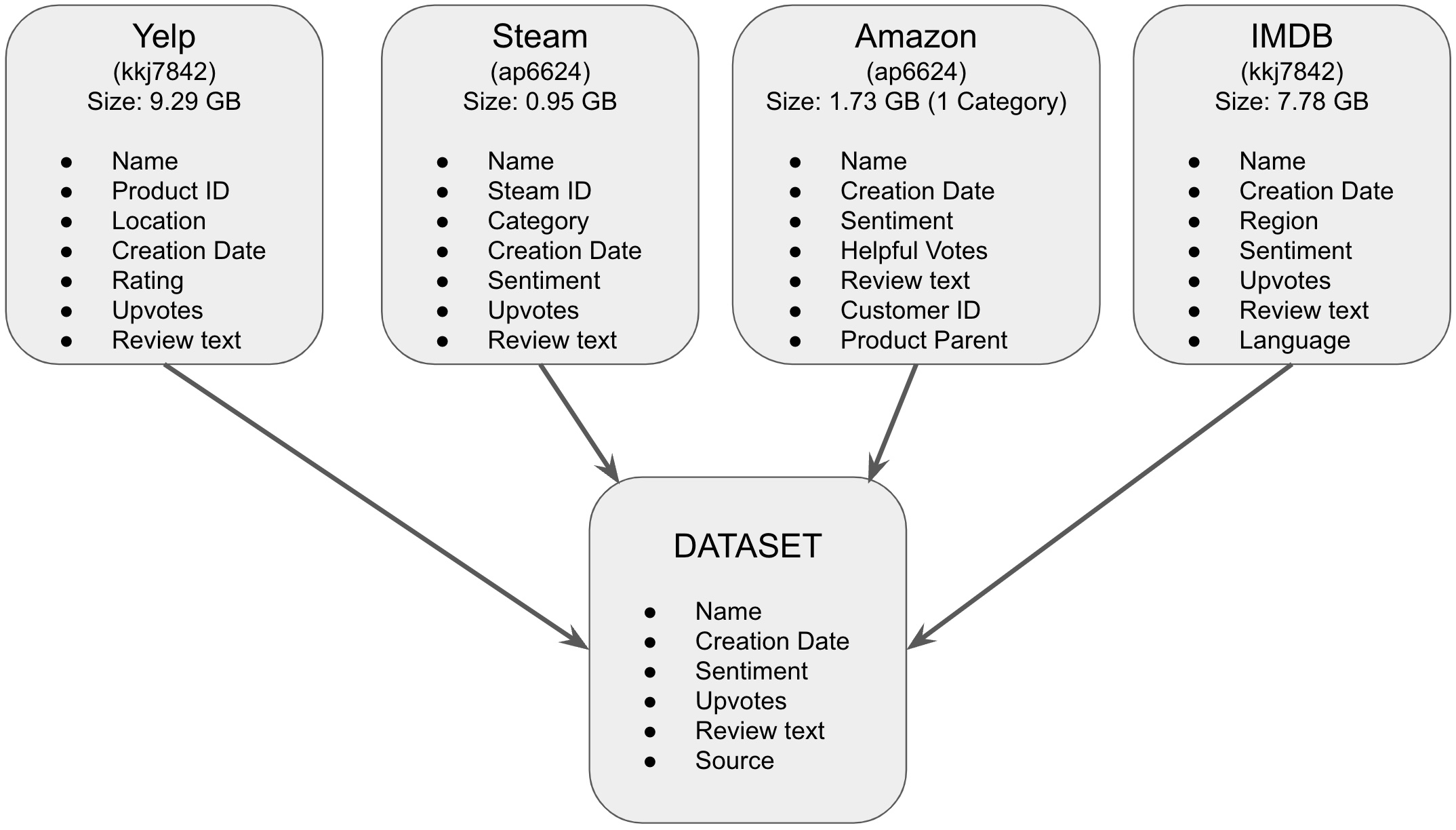}
  \caption{Datasets}
  \label{fig:datasets}
\end{figure*}

\subsection{Yelp}
The size of this dataset is 9.29 GB. The  \href{https://www.yelp.com/dataset/}{Yelp dataset}  is a collection of user reviews and ratings of businesses on the Yelp website. It contains information such as the text of the review, the date it was posted, the rating given by the user, and the user's location. The dataset is often used for research and analysis of consumer behavior and sentiment. It can be used to study trends in customer satisfaction, identify common complaints and compliments, and understand how different businesses are perceived by their customers. The dataset is also useful for training machine learning models to classify reviews and predict ratings.
\subsection{Steam}
\href{https://store.steampowered.com/}{Steam} is a popular platform for purchasing and playing PC games. Over the years, it has amassed a large number of reviews from its users. This dataset includes this review data, which can be used to analyze factors such as game satisfaction and dissatisfaction, genre popularity, and changes in sentiment over time. Note that this dataset was recently removed from Kaggle and the version used in this analysis was downloaded locally.

The size of this \href{https://www.kaggle.com/datasets/zeeenb/steamreviewsnlp}{dataset} is 950 MB.
\subsection{Amazon}
The Amazon customer review dataset is a valuable resource for understanding consumer opinions and experiences with products sold on \href{https://www.amazon.com/}{Amazon}. It includes millions of reviews written by Amazon customers, providing detailed information about the perceived quality of the products. The dataset can be used to analyze factors such as product popularity and satisfaction, as well as how the perception of products varies across geographical regions \cite{amazon_analysis}. This information can be useful for businesses looking to improve their products and customer experiences. \\
For the project, we have used the \href{https://www.kaggle.com/datasets/cynthiarempel/amazon-us-customer-reviews-dataset}{Amazon Electronics E-Commerce} subset and the size of this dataset is 1.73 GB.

\subsection{IMDb}
The size of this dataset is 7.78 GB. IMDb is a website that allows users to rate and review movies, TV shows, and other content. The \href{https://www.kaggle.com/datasets/ebiswas/imdb-review-dataset}{IMDb dataset} is a collection of user-generated reviews, ratings, and other metadata for a large number of films and TV shows. The dataset includes information such as the title of the film or show, the date of its release, the names of the cast and crew, and the user-generated review and rating. The dataset may also include other metadata, such as the number of times a review has been liked or disliked, or the number of users who have rated a particular film or show. This dataset is commonly used in research and analysis of user-generated content and can be used to study trends in user behavior, the relationship between ratings and reviews, and other aspects of the film and television industry.

\section{Design and Architecture}
\begin{figure}[h]
  \includegraphics[width=\linewidth]{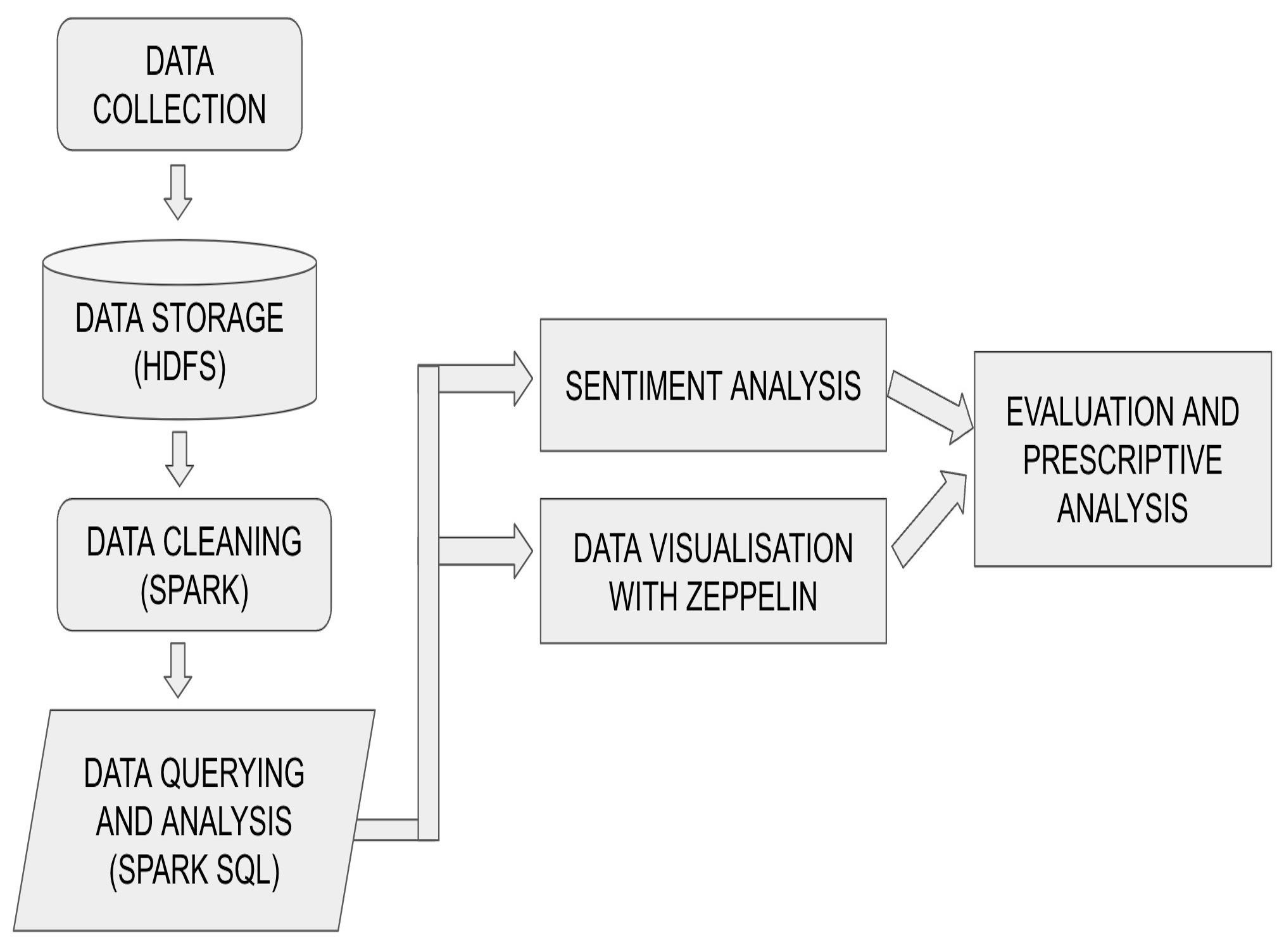}
  \caption{Data Pipeline}
  \label{fig:pipeline}
\end{figure}

In this project, we follow the data pipeline shown in Figure~\ref{fig:pipeline}. Each step is detailed in the following subsections.
\begin{figure*}[!]
  \includegraphics[width=\textwidth]{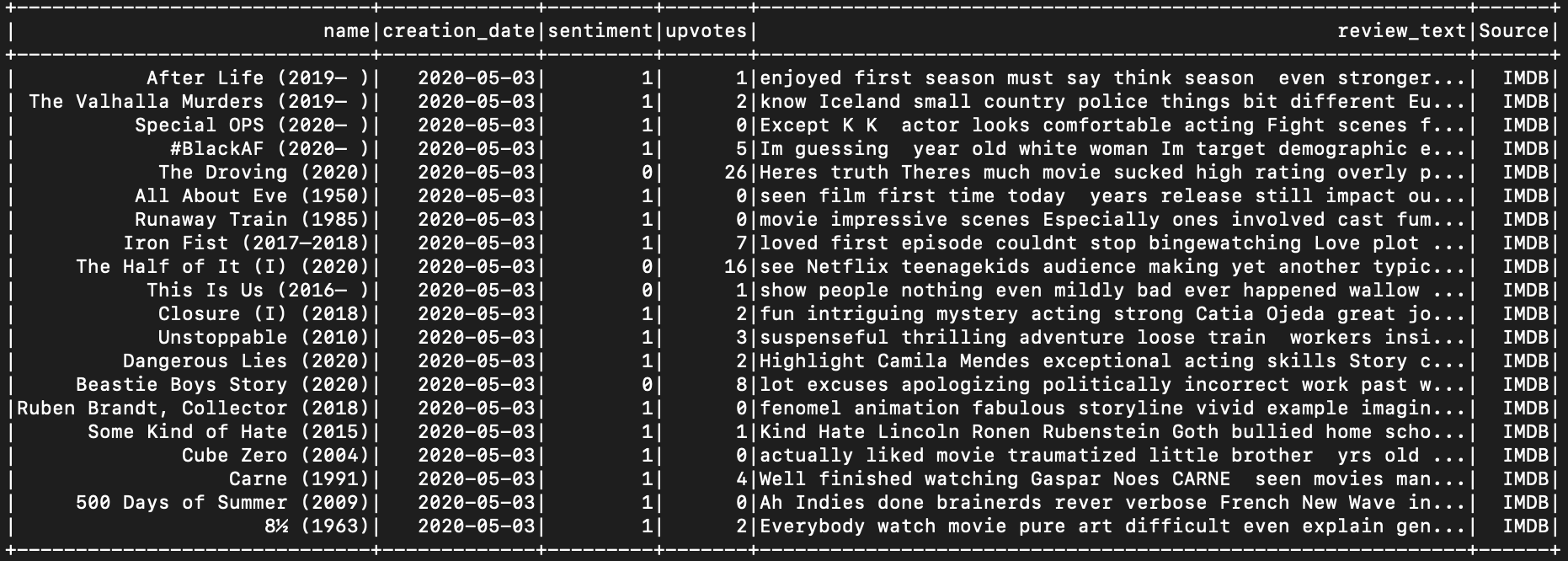}
  \caption{Screenshot of clean data}
  \label{fig:clean_data}
\end{figure*}
\subsection{Data Storage using HDFS}
In the beginning, we collected multiple consumer reviews from 4 datasets (mentioned in \hyperref[sec:dataset]{Section 4}) using the Kaggle API. \\
The datasets we obtained were in good condition, but we needed to do some additional cleaning and filtering to make them suitable for our task and analysis. We stored all the data (which was around 20 GB in size) on HDFS, a system that is designed for high-speed access to large amounts of data. We used Spark for cleaning the data.
\subsection{Data cleaning using Spark}
As there were 4 different datasets, each with its features and data points, we could not directly merge these datasets (as shown in Figure~\ref{fig:datasets}). After evaluating each of the datasets, we decided to include the following columns:
\begin{itemize}
    \item Name
    \item Creation Date (in yyyy-MM-dd format)
    \item Sentiment (0 or 1)
    \item Upvotes
    \item Review Text
    \item Source
\end{itemize}

The next step was to clean the data. All our datasets are individual files stored in a CSV format. We used Spark to clean the data and performed the following steps:
\begin{itemize}
    \item For all data, we trimmed the string of leading spaces and quotes.
    \item Filtered out the unwanted columns and rows containing null values and noisy data
    \item For the review text specifically, we performed the following cleaning steps:
    \begin{itemize}
        \item Removed non-alphabetic characters
        \item Removed digits
        \item Removed stopwords. Words that have little or no significance, especially when constructing meaningful features from the text are also known as stop words. Words like a, an, the, and so on are considered to be stopwords. This is a common step in Language Analysis and it usually improves performance.
    \end{itemize}
    \item In our datasets, we have different variations for sentiment labels. The Yelp and IMDB dataset have a 5-class sentiment (\texttt{very negative}, \texttt{negative}, \texttt{neutral}, \texttt{positive}, \texttt{very positive}), whereas the Steam dataset has just 2 classes, \texttt{positive} and \texttt{negative}. We decided to convert all the labels to just 2 classes, \texttt{positive} and \texttt{negative}.
    \item Review date was processed to get to the uniform form yyyy-mm-dd from varying formats.
    
\end{itemize}
The cleaning steps in this process are performed using Spark transformations and actions on RDD(s) (Resilient Distributed Dataset) representing the input CSV file. The transformation processes one row of the CSV input at a time, following the specified steps to clean the data. The cleaned data is then sent to an action, which writes the output to HDFS (as well as a table). This process uses Spark and RDDs to efficiently clean and process the data.

\subsection{Data Querying and Analysis using Spark SQL}

The Spark data-frames were then combined into a single dataframe and then were performed analysis on this data-frame using Spark SQL.

Spark SQL is a component of Apache Spark that allows users to process structured data, such as data stored in tables. Spark SQL provides a programming interface that allows users to express their queries and manipulations in a syntax that is similar to SQL, a widely-used language for querying and manipulating databases.  Figure~\ref{fig:clean_data} is a screenshot of the cleaned data from the dataset. 

\subsection{Data Visualization using Zeppelin}
Apache Zeppelin \cite{zeppelin} is a web-based notebook that enables data-driven, interactive data analytics. It provides support for a wide range of languages and technologies, like Spark (and Spark SQL) and Scala. With Zeppelin, users can create and share documents that contain code, visualizations, and text - similar to Jupyter Notebooks. This makes it a convenient and interactive environment for working with big data and performing data analytics and visualization. Zeppelin also supports collaboration, allowing multiple users to access and edit the same document.
\\
We created a union of our 4 individual tables to jointly analyze and run queries on the entire dataset. The graphs and charts are detailed in the next section.

\section{Results}

\subsection{Increase in reviews over the years}
As expected, one simple result we saw was that user engagement with social sites has increased significantly. This corresponds to the increase in the use of hand-held devices and readily available internet services in recent years. With the available stats, we also found that during the pandemic years 2020-2022, there was a spike in the count of reviews due to more indoor activities as compared to the pre-covid years.

\subsection{Year on Year Trends}
\begin{figure}[H]
  \includegraphics[width=\linewidth]{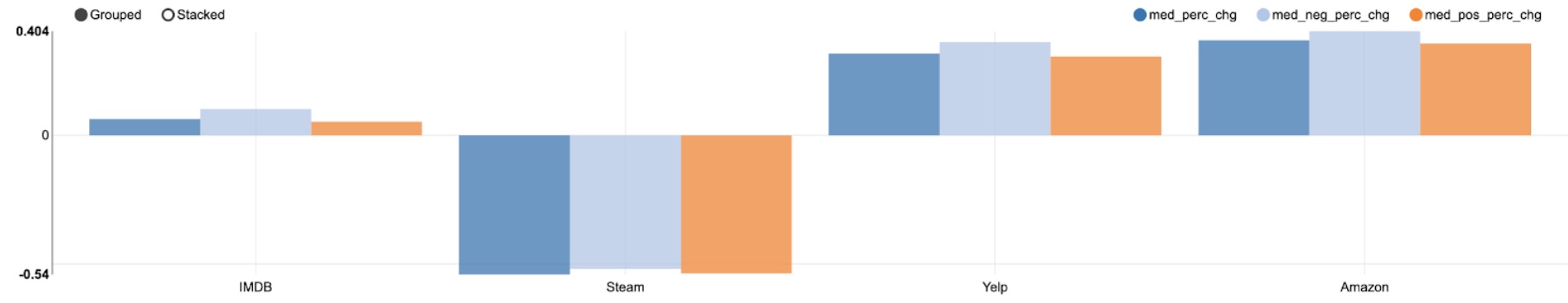}
  \caption{Year on Year Trends}
  \label{fig:YearOnYear}
\end{figure}

We see that the median percentage change of users engaging on IMDB, Yelp, and Amazon are all positive - this means that more and more users are using these sites and buying/using these products and services. 

This corresponds with an increase in personal devices, an increase in the number of movies \& products as well as more easily accessible to the internet in the past few years.
On the other hand, we see that Steam has a trend of decreasing the number of reviews on its site. 

This speaks to a few factors - more and more people have started playing mobile and console games as compared to PC games in recent years. The declining popularity of Steam as a platform also adds to this. \cite{steam_genre}

One interesting observation on this view of data was that the increase in negative reviews was higher than the increase in positive reviews. This indicated that people have become more critical of services over time especially as the internet has grown.

\subsection{User Engagement on a Weekday Basis}
\begin{figure}[H]
  \includegraphics[width=\linewidth]{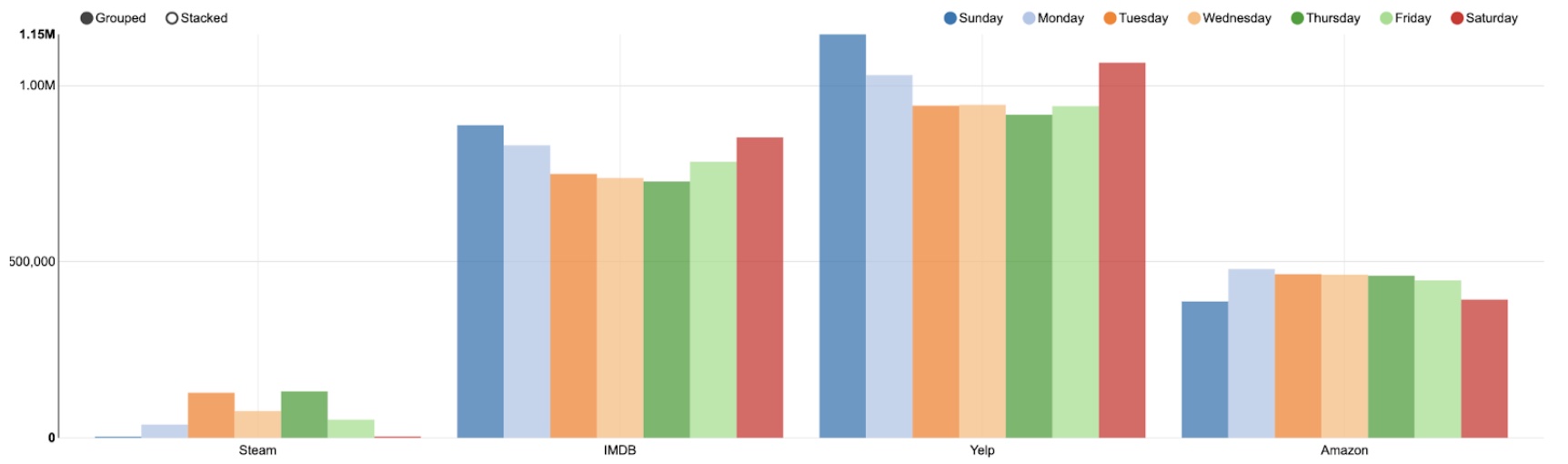}
  \caption{Reviews Per Weekday}
  \label{fig:weekday}
\end{figure}

Looking at the data per weekday in \ref{fig:weekday}, we see that there is a significantly higher number of user engagement on IMDB and Yelp on Saturdays, Sundays, and Mondays. This is intuitively correlated with the fact that most users/people go out of their houses over the weekend (Fri-Sun) and then post reviews which lag by a day. 
On the other hand, Amazon sees a decrease in the number of reviews posted on the weekend which can be explained by the fact that most deliveries take place on Business Days as well as more people spending time outside of their houses. 

\subsection{User Engagement on a Monthly Basis}
\begin{figure}[H]
  \includegraphics[width=\linewidth]{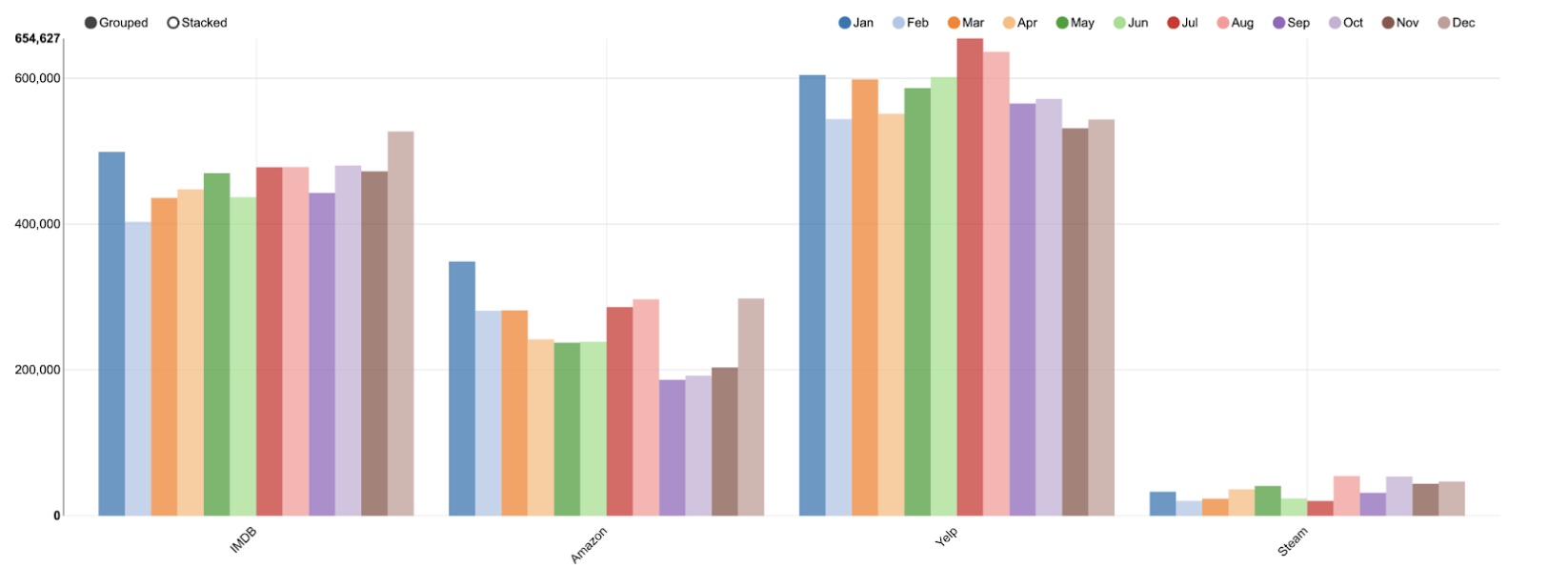}
  \caption{Reviews Per Month}
  \label{fig:monthly}
\end{figure}

\begin{figure}[H]
  \includegraphics[width=\linewidth]{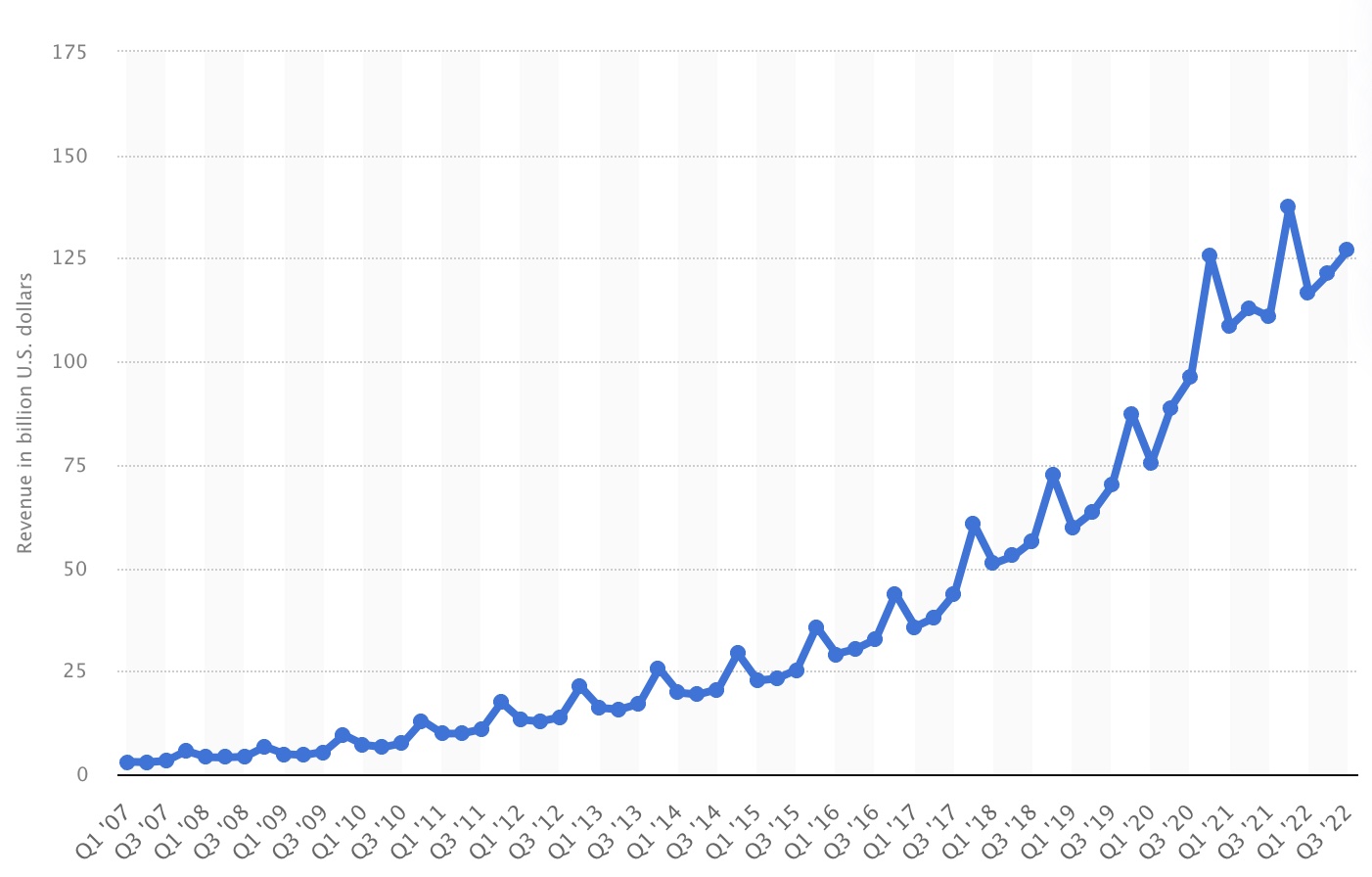}
  \caption{Amazon Revenue/Quarter \cite{amazon_rev_quarter}}
  \label{fig:amazon_revenue}
\end{figure}

From \ref{fig:monthly}, one clear trend we see on Amazon and IMDB is that user engagement with the specified social sites increases significantly during the winter period which is intuitive due to the holiday and sale season.
This can be corroborated by the increase in sales reported by Amazon \cite{amazon_rev_quarter} (Figure~\ref{fig:amazon_revenue}). 


We also see a significant increase in the reviews on Yelp in addition to the other two sites during the Summer time which coincides with people going out of their houses as well as increased spending. 

For Steam, we noticed the increase in reviews just after the summer as Aug and Oct/Nov periods are usually the prime months for new game releases and the winter season due to increased indoor activity. We can still see a small spike at end of the Summer months which can be attributed to the "End of Summer sales" by Steam \cite{steam_summer}.

\subsection{Correlating Upvotes and Review Length}
An analysis of the relationship between review length and upvotes reveals an interesting result. The number of reviews decreases exponentially with the number of characters in the review. However, longer reviews were found to be significantly more helpful than shorter reviews. 
\begin{figure}[H]
  \includegraphics[width=\linewidth]{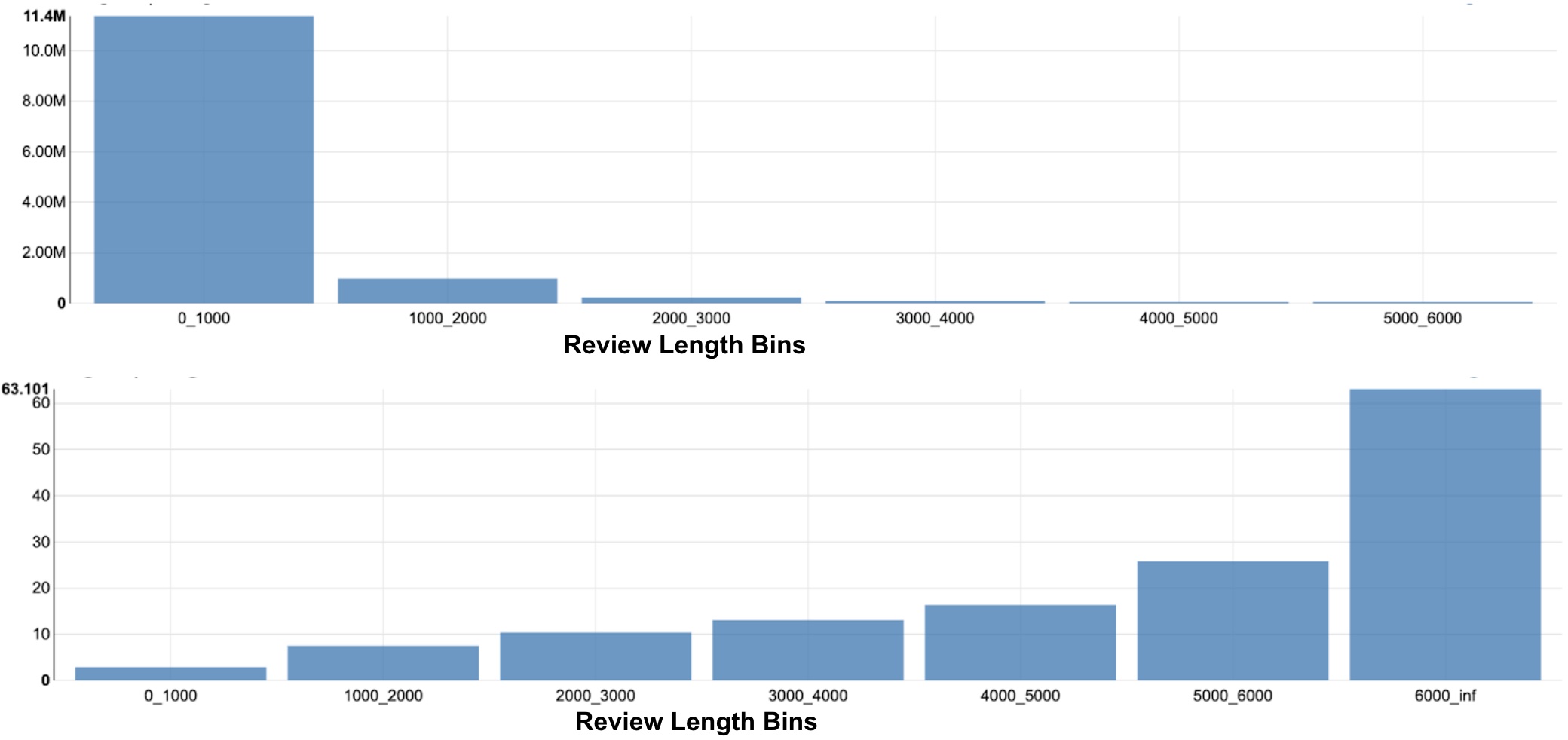}
  \caption{Correlating upvotes and review length}
  \label{fig:upvotes_rev_len}
\end{figure}
In general, it can be concluded that longer reviews are more informational and provide better insights into the product being reviewed. But unfortunately, most reviewers do not put in the effort to write such long and detailed reviews.

\subsection{Correlating Sentiment against Upvotes and Review Length}
\begin{figure}[H]
  \includegraphics[width=\linewidth]{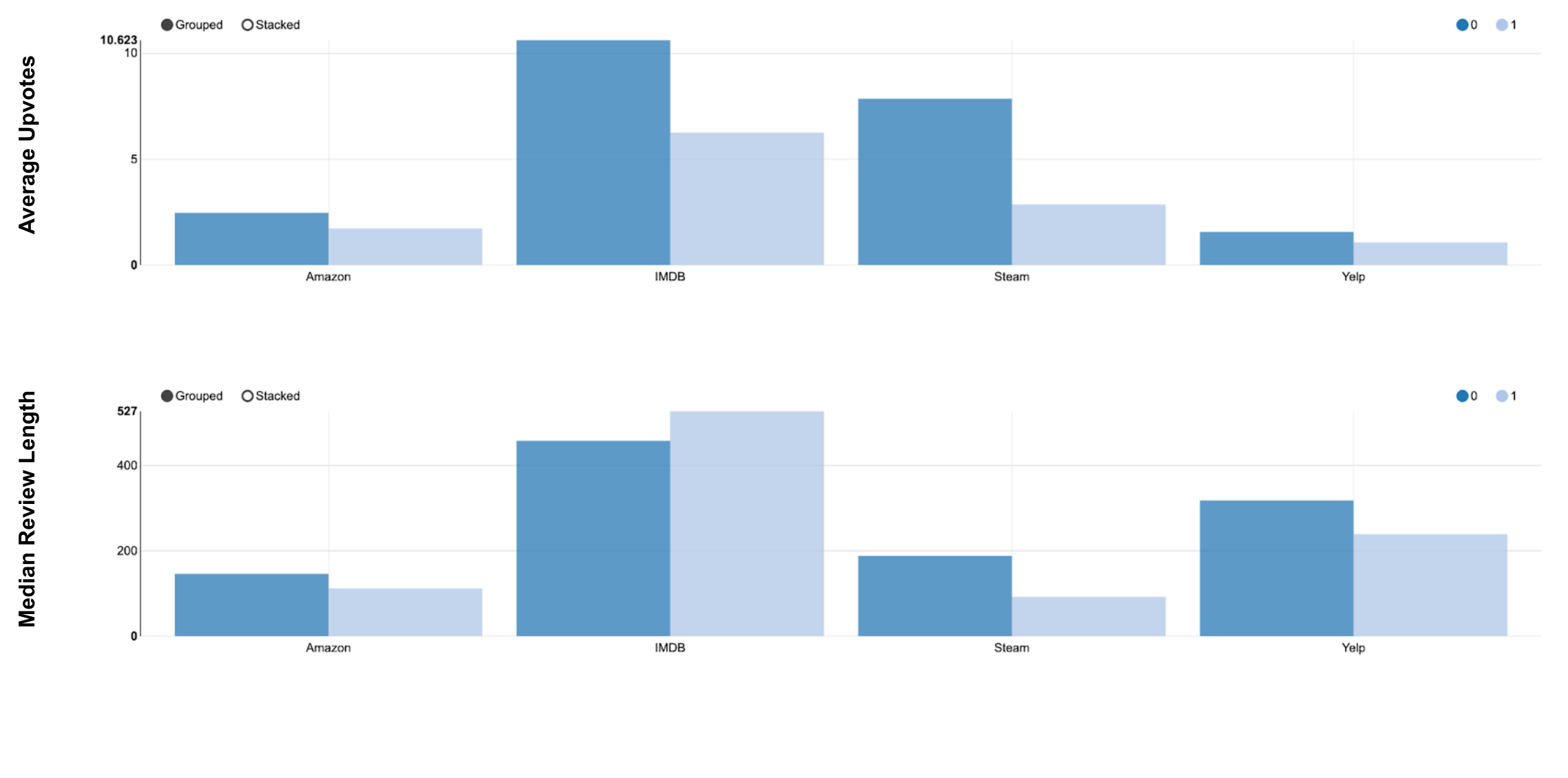}
  \caption{Correlating sentiment against upvotes and review length}
  \label{fig:sentiment_vs_rev_up}
\end{figure}

We correlate sentiment against upvotes and review lengths in Figure~\ref{fig:sentiment_vs_rev_up}.
For all datasets except IMDb, the negative reviews are longer than the positive reviews. 
IMDb is an outlier with longer positive reviews than negative reviews.
We can conclude that reviewers with a negative opinion tend to be more expressive when writing reviews. IMDb’s outliers can be alluded to reviewers being art enthusiasts who tend to be more eloquent.
For all datasets, average upvotes for negative reviews are higher than average upvotes for positive reviews.
This signifies that customers value negative reviews more than positive ones since reviews tend to shape their decision to purchase the product. (or visit the establishment on Yelp)

\section{Conclusion}
To summarize, we used various big data tools, such as Hadoop, Spark, Spark SQL, and Zeppelin, to perform sentiment, language, and user engagement analysis on social multimedia data. Our analysis showed that consumer reviews are important, and sentiment analysis is a growing area of research in text mining. In conclusion, this project demonstrated the value of using big data tools for analyzing social multimedia data.

\section{Future Scope}
We plan to extend this project to include real-time data and to analyze it daily. We feel that by increasing the timeframe over which we do our analysis, we will get stronger results to be able to back up our claims more conclusively. Additionally, we plan to add more parameters to our analysis. Finally, we aim to create a dashboard that makes it easy to visualize and distribute sentiment analysis for products.

\Urlmuskip=0mu plus 1mu\relax
\bibliographystyle{unsrt}
\bibliography{bibflile}

\end{document}